\documentclass[conference]{IEEEtran}
\IEEEoverridecommandlockouts
\usepackage{cite}
\usepackage{amsmath,amssymb,amsfonts}
\usepackage{algorithmic}
\usepackage{graphicx}
\usepackage{textcomp}
\usepackage{xcolor}
\usepackage{comment}
\usepackage{subfigure}
\usepackage{algorithmic}
\usepackage[ruled,vlined]{algorithm2e}
\usepackage{adjustbox}

\def\BibTeX{{\rm B\kern-.05em{\sc i\kern-.025em b}\kern-.08em
    T\kern-.1667em\lower.7ex\hbox{E}\kern-.125emX}}
\begin{document}
\title{Accelerating and Compressing Deep Neural Networks for Massive MIMO CSI Feedback
}

\author{\IEEEauthorblockN{Omar Erak and Hatem Abou-Zeid}

\IEEEauthorblockA{{Department of Electrical and Software Engineering}, 
{University of Calgary},
Calgary, Canada\\ \{omar.erak, hatem.abouzeid\}@ucalgary.ca }
}
\maketitle

\begin{abstract}
The recent advances in machine learning and deep neural networks have made them attractive candidates for wireless communications functions such as channel estimation, decoding, and downlink channel state information (CSI) compression. However, most of these neural networks are large and inefficient making it a barrier for deployment in practical wireless systems that require low-latency and low memory footprints for individual network functions. 
To mitigate these limitations, we propose accelerated and compressed efficient neural networks for massive MIMO CSI feedback. Specifically, we have thoroughly investigated the adoption of network pruning, post-training dynamic range quantization, and weight clustering to optimize CSI feedback compression for massive MIMO systems. Furthermore, we have deployed the proposed model compression techniques on commodity hardware and demonstrated that in order to achieve inference gains, specialized libraries that accelerate computations for sparse neural networks
are required. 
Our findings indicate that there is remarkable value in applying these model compression techniques and the proposed joint pruning and quantization approach reduced model size by 86.5\% and inference time by 76.2\% with minimal impact to model accuracy. These compression methods are crucial to pave the way for practical adoption and deployments of deep learning-based techniques in commercial wireless systems.
\end{abstract}


\section{Introduction}
In recent years, machine learning (ML) has proven to be a fast, reliable, and robust alternative for solving complex problems in different fields, such as computer vision\cite{b1}, autonomous driving\cite{b2} and medical diagnosis\cite{b3}. As a result, there is a growing interest to explore the potential applications of machine learning in wireless communications to improve performance and reliability. 
For example, \cite{b4} have demonstrated promising performance for different coding schemes such as low-density parity-check (LDPC) and BCH codes using a graph neural network (GNN)-based architecture for channel decoding. The work in \cite{b5} has shown that it is possible to use a neural network (NN)  for the synchronization of narrowband physical random-access channel (NPRACH) for uplink narrowband internet of things (NB-IoT). Machine learning has also been used for downlink applications. CsiNet was developed using deep learning and has proven to be a successful channel state information (CSI) sensing and recovery mechanism that outperformed existing compressive sensing (CS)-based methods \cite{b6}.

The performance and reliability improvements that have been achieved as a result of applying machine learning techniques in wireless communications continues to drive further research forward. However, these improvements come with a hefty cost that is often neglected. These machine learning solutions are usually computationally demanding and they require a large amount of resources, including memory, CPU, and energy to train and deploy the models, and this has significantly hindered their practical deployment.

Recently, there has been a growing focus on developing tiny and efficient machine learning models that are small enough to run on embedded systems with limited resources. By applying pruning, trained quantization and huffman coding \cite{b7} were able to compress AlexNet, LeNet and VGG-16 by 35$\times$, 39$\times$ and 49$\times$ respectively with no loss in accuracy. More recently, a convolutional neural network (CNN) architecture called SqueezeNet was developed and it was able to achieve the same accuracy AlexNet achieved on ImageNet whilst being 510$\times$ smaller in size\cite{b8}. These examples showcase the tremendous success of developing tiny, efficient neural networks in computer vision. However, there has been limited work accelerating and compressing deep neural networks in wireless communications.

Our work aims to demonstrate the potential and importance of accelerating and compressing deep learning based wireless communications solutions. Developing efficient models will pave the way for practical adoption and deployments in commercial systems. The main contributions of this paper are:

\begin{itemize}
\item We have investigated the adoption of three common model compression techniques, namely: pruning, post-training quantization and weight clustering to optimize CsiNet, a CNN developed for massive MIMO CSI feedback \cite{b6}. We have demonstrated how each model compression technique individually affects the model storage, prediction accuracy, and inference time - as well as the results of combining multiple techniques together.
\item A thorough investigation of the degree of neural network pruning and different quantization levels has been conducted, and new insights on the resulting performances and trade-offs are drawn.
\item We have deployed the proposed model compression techniques on a Raspberry Pi 4 and demonstrated that to achieve inference gains, specialized libraries that accelerate computations for sparse neural networks are required. To the best of our knowledge, this is the first study to deploy CsiNet compressed models on commodity hardware, use acceleration libraries, and measure inference times. 
\item The proposed compression method that combines network pruning and quantization achieved an average of 86.5\% reduction in model size and reduced inference time by 76.2\% compared to the original CsiNet models with minimal impact to model accuracy.
\end{itemize}

The results of this paper highlight the significant gains model compression techniques can have for wireless communications deep learning models. All the code to reproduce our results and conduct further research to other CSI compression networks, and other wireless communications models will be made available upon acceptance.

The rest of this paper is organized as follows. In Section \ref{relatedwork} we provide a summary of related research. Following that, Section \ref{background} presents the system model of CsiNet \cite{b6} and our proposed model compression techniques for CsiNet. In Section \ref{setup} we describe our experimental setup and evaluation metrics, and Section \ref{results} discusses the results of our findings. Finally, we discuss conclusions and   future directions in Section \ref{conclusion}.



\section{Related Work} \label{relatedwork}


A limited number of studies have investigated and researched the development of the low-complexity deep learning models in the context of ML for wireless communications. Recently, \cite{b10} developed a tiny ML approach for channel estimation and signal detection. Their approach splits each large layer in a deep neural network into three smaller sub-layers, and they were able to achieve 4.5$\times$ reduction in model storage. In \cite{b11}, the authors have explored the applications of some model compression techniques and demonstrated their potential for signal detection in MIMO communications and CSI feedback compression.  
The authors of \cite{b12} and \cite{b13} developed lightweight neural networks for CSI feedback in massive MIMO. In \cite{b12}, a new lightweight network called ShuffleCsiNet that achieved around 5$\times$ reduction in parameter number compared to ConvCsiNet is proposed, and in \cite{b13} the authors reduce the complexity of existing networks by 25.50\%-43.46\%  using knowledge distillation, with minimal impacts to accuracy.

Despite the promising results achieved, a comprehensive understanding of neural network pruning, quantization, and weight clustering on deep learning compression models for massive MIMO CSI feedback has not been conducted. To this end, in this paper we have applied these standard model compression techniques and analyzed both their individual and combined performances to a deep learning network for CSI feedback compression. We have measured the key  quantitative metrics of model storage, prediction accuracy, and inference time. Different from prior work, we measure real-world \emph{inference time} on a Raspberry Pi 4 after deploying the compressed deep learning models on the hardware. We then show that specialized libraries that can accelerate the computations for sparse neural networks are required to realize the inference gains that are cited in literature as theoretical reductions in computational complexity. Our results indicate that without such accelerators there may be limited real-world gains in inference time.

\section{Accelerated and Compressed Deep Learning Models for CSI Feedback} 
\label{background}

\subsection{System Model of CsiNet}

\begin{figure*}
\centering
		\includegraphics[keepaspectratio, width=0.89\textwidth, height=3in]{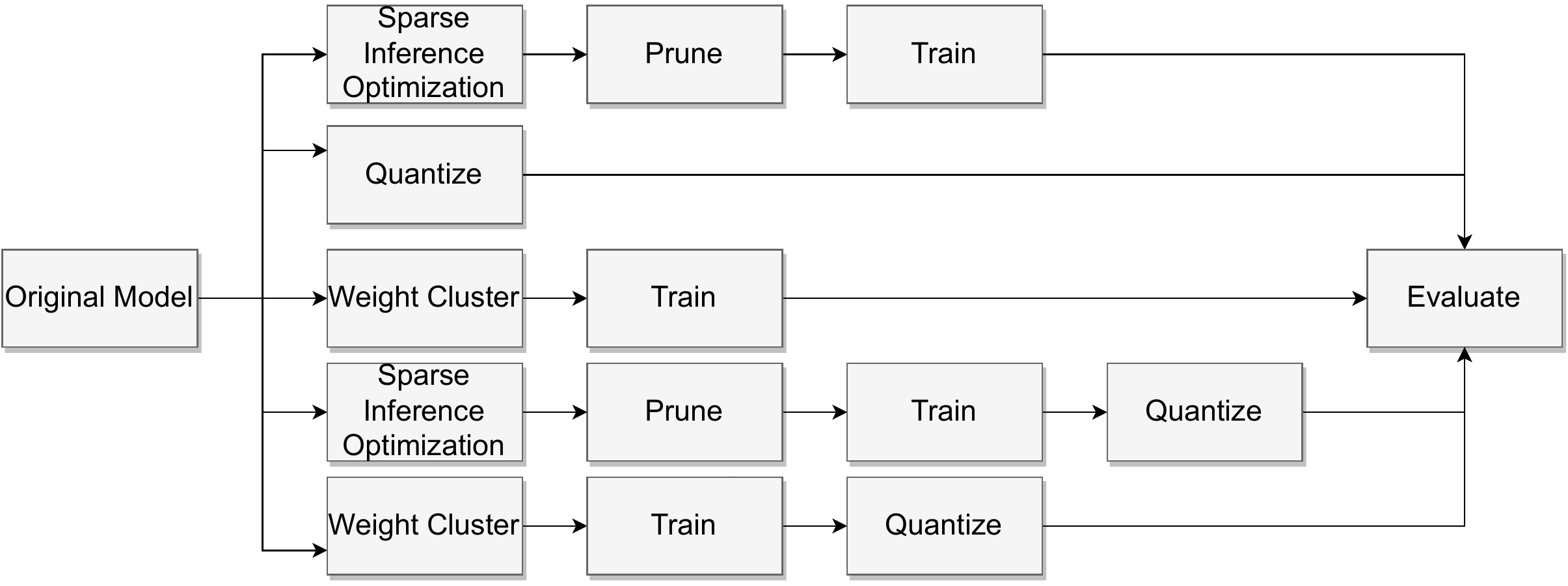}
		\caption{Process of Applying Different Model Compression Techniques}
		\label{Workflow}
\end{figure*}

CsiNet is a CNN that is used for channel state information (CSI) sensing and recovery mechanism. This model uses an encoder (sensing) and decoder (recovery) to compress CSI at the user equipment (UE) and reconstruct it at the base station (BS). CsiNet outperformed traditional compressive sensing algorithms at all compression ratios \cite{b6}.

When developing CsiNet, the authors of \cite{b6} considered a massive single-cell downlink MIMO system with a single receiver antenna at the UE. The number of transmitting atennas at the BS is $N_{t}\gg1$. This system operates in OFDM over $\hat{N_{c}}$ subcarriers. 

Let $\hat{\textbf{H}}$ be the CSI matrix in the spatial frequency domain acquired at the UE. To reduce the significant impact of feedback, \cite{b6} proposed converting $\hat{\textbf{H}}$ into an angular delay domain matrix ${\textbf{H}}$ using a 2D discrete Fourier transform (DFT) as follows:

\begin{equation}
    {\textbf{H}} = {\textbf{F$_d$}}\hat{\textbf{H}}{\textbf{F$^H_a$}}
\end{equation}

where {\textbf{F$_d$}} is a $\hat{N_{c}}$ $\times$ $\hat{N_{c}}$ DFT matrix  and {\textbf{F$_a$}} is a $N_{t}$ $\times$ $N_{t}$ DFT matrix. The total number of feedback parameters is given by $N$ = $N_{t}$ $\times$ $\hat{N_{c}}$.

CsiNet implements an encoder that takes the angular delay domain matrix ${\textbf{H}}$ as input applies the encoding function ${\textbf{f}}$ and returns an $M$ dimensional encoded vector $\textbf{s}$, where $M < N$ as shown below:

\begin{equation}
    {\textbf{s}} = {\textbf{f(\textbf{H})}}
\end{equation}

The compression ratio \(\gamma\) for the encoded vector is given by $M/N$. 
In this paper, we will evaluate the performance of CsiNet, and the proposed efficient and compressed CsiNet networks for various compression ratios.
CsiNet also implements a decoder at the BS that takes $\textbf{s}$ as the input and ideally returns the original angular delay domain matrix ${\textbf{H}}$ as follows:

\begin{equation}
    {\textbf{H}} = {\textbf{f$^{-1}$(\textbf{s})}}
\end{equation}

Let $\Tilde{\textbf{H}}$ be the recovered channel from CsiNet. To evaluate the performance of the CSI reconstruction the normalized mean square error (NMSE) is used as follows:

\begin{equation}
    \label{eq:nmse}
    \textbf{NMSE} = \mathbb{E}\left\{\frac{||\Tilde{\textbf{H}} - \textbf{H}||^2_2}{||\textbf{H}||^2_2}\right\}
\end{equation}

Another metric used by \cite{b6} to evaluate the performance of CsiNet is the cosine similarity \(\rho\) to measure the quality of the beamforming vector ${\textbf{v}}_n$. \(\rho\) and ${\textbf{v}}_n$ are given as follows: 

\begin{equation}
    \textbf{v}_n= \frac{\textbf{h}_n}{||\textbf{h}_n||_2}
\end{equation}

\begin{equation}
    \label{eq:rho}
    \rho  = \mathbb{E}\left\{\frac{1}{\Tilde{N_c}}\sum_{n=1}^{\Tilde{N_c}}\frac{|\textbf{h}^H_n\hat{\textbf{h}}_n|}
    {||\textbf{h}_n||_2||\hat{\textbf{h}_n||_2}}\right\}
\end{equation}

where $\textbf{h}_n$ is the reconstructed channel vector and $\hat{\textbf{h}_n}$ is the original channel vector of the $n$th subcarrier.

The training and testing samples used in this paper are equivalent to the ones used by \cite{b6} to ensure a fair comparison is made. The channel matrices for two environments are created using the COST 2100 model \cite{b14.5}. The first environment is the indoor picocellular at the 5.3GHz band and the second environment is the outdoor rural environment at the 300 MHz band. All parameters follow their default setting in \cite{b14.5}.

\subsection{Compression Techniques}
In this work, we apply three common model compression techniques individually and then we combine post-training quantization with pruning and with weight clustering. Figure ~\ref{Workflow} summarizes the workflow used to apply the different techniques and evaluate their overall performance.

\subsubsection{\textbf{Magnitude-Based Weight Pruning}} Pruning is a model compression technique used to remove parameters from a large, dense network, yielding a sparse subnetwork that is smaller in size and more computationally efficient compared to the original network. There are different pruning methods and workflows, in this paper we apply magnitude-based weight pruning which works by ranking connections in a network according to the absolute values of their weights, i.e how much they contribute to the overall output. A target sparsity ratio is selected and low-weight connections that have minimal impact are set to 0 and then removed to achieve the targeted sparsity ratio. The resultant sparse model is then retrained so new weights can be determined in order to maintain similar accuracy \cite{b9} \cite{b9.5}.

Firstly, we restructure the original model to comply with the constrains of XNNPack sparse inference for CNN models \cite{b15} \cite{b16}. XNNPack is a highly optimized library of neural network inference operators that allows us to improve the inference performance for the sparse pruned model. Common nueral network operators that are used in the original CsiNet model and are supported by XNNPack include: 2D Convolution, Leaky ReLU and sigmoid \cite{b6}\cite{b16}. Then, we apply magnitude based weight pruning to the restructured CsiNet model, we apply pruning to the dense layers and we avoid pruning the first and final layers so that the model's accuracy is not significantly impacted. We also use the PruneForLatencyOnXNNPack pruning policy which ensures that only parts of the model that are compliant with XNNPack sparse inference are pruned \cite{b9}\cite{b15} \cite{b16}. Finally, we fine-tune the pruned model by retraining it. Algorithm ~\ref{alg1} presents the pseudo-code for magnitude-based weight pruning of dense layers.

\subsubsection{\textbf{Post-Training Quantization}} This is a model compression technique that works by storing weights and activations with a lower precision, for example, weights can be stored as 16 bit floating point values or as 8 bit integers instead of the full precision of 32 bit floating point values. This would allow for around $2\times$ to $4\times$ reduction in model size. Furthermore, hardware support for 8 bit integer computations is typically faster compared to 32 bit floating point values computations, which helps improve inference time.

We apply quantization to a trained model and we evaluate the quantized model without retraining. 
We apply dynamic range quantization which statically quantizes only the weights and activations from floating point to integer and performs computations with 8-bit weights and activations \cite{b9}. The above method is unchanged when we combine post-training quantization with other techniques.

\subsubsection{\textbf{Weight Clustering}} This is a model compression technique that works by grouping $n$ weights $W = \{w_1,w_2...,w_n\}$ of each layer in a network into $k$ clusters $C = \{c_1, c_2,...,c_k\}$ where $n\gg k$. The clusters' centroids are then initialized using a centroid initialization method such as linear, density, random and kmeans++ initialization. The clusters' centroid value is then shared for all the weights belonging to the cluster. The model is then fine-tuned by retraining for less epochs in order to maintain a high accuracy \cite{b7} \cite{b9}.

We apply weight clustering to a trained model. We only apply weight clustering to the dense layers and we avoid weight clustering the first and final layers so that the model's accuracy is not significantly impacted. Finally, we fine-tune the weight clustered model by retraining it.

\subsubsection{\textbf{Combining Techniques}} In this paper we combine dynamic range quantization with pruning and dynamic range quantization with weight clustering. Figure ~\ref{Workflow} summarizes the workflow used to combine these compression techniques. For pruning and weight clustering, the models needs to be retrained so that they are fine-tuned which helps maintain a high accuracy. Therefore we apply quantization in the final step once the models have been trained so that the new retrained weights are quantized. Algorithm ~\ref{alg1} presents the pseudo-code for combining pruning with quantization.

\begin{algorithm}[t]
\caption{CsiNet Magnitude-Based Weight Pruning and Quantization}\label{alg1}
\textbf{Input:} \textit{model:} CsiNet Model \textit{ratio:} Sparsity Ratio \textit{level:} Quantization Level

\textbf{Output:} \textit{final model:} Pruned Quantized Model

    \SetKwFunction{FMain}{PruneQuantize}
    \SetKwProg{Fn}{Function}{:}{}
    \Fn{\FMain{$model$, $ratio$, $level$}}{
        $model$ $\Leftarrow$ \textit{SparseModelOptimization(model)} 

        \ForEach{$ layer \in model.layers $}
        {\If{$ layer == Dense $}
            {$T$ $\Leftarrow$ \textit{TotalNumberOfWeights(layer.weights)}
            
        $S$ $\Leftarrow$ \textit{Sort(layer.weights)}
        
        $min$ $\Leftarrow$ $S[integer(T*ratio)]$
            
        \ForEach{$ weight \in layer.weights $}
        {
            {\If{$ weight < min $}
            {$weight$ $\Leftarrow$ 0}
            }
        }
        }
        }
        $model$ $\Leftarrow$ \textit{FineTuneTraining(model)}
                
        \textit{final model} $\Leftarrow$ \textit{Quantization(model, level)}
        
        \textbf{return} \textit{final model}
        }
\end{algorithm}

\section{Experimental Setup}\label{setup}

\subsection{Hardware and Software Platforms}
The results in this paper were obtained using the following hardware and software platforms:

\textbf{Hardware} The system used to conduct the experiments is a Raspberry Pi 4. It has a 64-bit quad-core ARM Cortex-A72 (BCM2711) CPU running at 1.5GHz. This system has 4GB of LPDDR4-3200 SDRAM.

\textbf{Software} The system used to conduct the experiments runs the Raspberry Pi OS, Kernel version: 5.15
Debian version: 11 (bullseye). We use Python (v3.9.2) and TensorFlow (2.10.0). We use the TensorFlow Lite interpreter to perform inference, and benchmark using the TensorFlow Lite benchmark tool.

\subsection{Evaluation Metrics}
To evaluate the effectiveness of pruning, post-training quantization and weight clustering on the CsiNet models we use the following metrics:

\textbf{Model Size} \emph{(lower is better)} This is the model binary size once it is saved on disk. This metric is measured in megabytes.

\textbf{Inference Time} \emph{(lower is better)} This is the time it takes the model to process the input and produce a valid output. This metric is measured in microseconds.

\textbf{Normalized Mean Square Error} \emph{(lower is better)} This is the metric described by Equation \ref{eq:nmse} to evaluate the performance of the model. It is the normalized difference between the recovered channel and the original channel.

\textbf{Cosine Similarity} \emph{(higher is better)} This is the metric described by Equation \ref{eq:rho} and is also used to evaluate the performance of the model. This metric measures the quality of the beamforming vector and will be referred to as  \(\rho\).

\begin{table*}[!htb]
\centering\small
\caption{Individual Model Compression Techniques on CsiNet}
\label{table:CsiNetResults1}
\begin{adjustbox}{width=0.9\textwidth}
\begin{tabular}{|l|l|l|l|l|l|l|l|}
\hline
\textbf{\(\gamma\)} & \textbf{Model}                                                                                & \textbf{Size (MB)}                                & \textbf{Inference (\(\mu\)s)}                           & \textbf{Indoor NMSE}                              & \textbf{Indoor \(\rho\)}                                 & \textbf{Outdoor NMSE}                             & \textbf{Outdoor \(\rho\)}                                \\ \hline
1/4        & \begin{tabular}[c]{@{}l@{}}CsiNet\_Original\\ CsiNet\_Quantized\\ CsiNet\_Pruned\\ CsiNet\_Weight\_Clustered\end{tabular} & \begin{tabular}[c]{@{}l@{}}8.037\\2.031\\ 4.423\\ 1.519\end{tabular} & \begin{tabular}[c]{@{}l@{}}8315.44\\ 6434.42\\ 3792.93\\ 8097.44\end{tabular} & \begin{tabular}[c]{@{}l@{}}-17.36\\ -17.35\\ -17.51\\ -11.69\end{tabular} & \begin{tabular}[c]{@{}l@{}}0.99\\ 0.99\\ 0.99\\ 0.96\end{tabular} & \begin{tabular}[c]{@{}l@{}}-8.75\\ -8.72\\ -8.70\\ -9.82\end{tabular} & \begin{tabular}[c]{@{}l@{}}0.91\\ 0.91\\ 0.91\\ 0.93\end{tabular} \\ \hline
1/16       & \begin{tabular}[c]{@{}l@{}}CsiNet\_Original\\ CsiNet\_Quantized\\ CsiNet\_Pruned\\ CsiNet\_Weight\_Clustered\end{tabular} & \begin{tabular}[c]{@{}l@{}}2.037\\ 0.531\\ 1.137\\ 0.384\end{tabular} & \begin{tabular}[c]{@{}l@{}}6381.05\\5921.32\\ 2130.89\\ 6310.41\end{tabular} & \begin{tabular}[c]{@{}l@{}}-8.65\\ -8.65\\ -8.67\\ -7.39\end{tabular} & \begin{tabular}[c]{@{}l@{}}0.93\\ 0.93\\ 0.93\\ 0.91\end{tabular} & \begin{tabular}[c]{@{}l@{}}-4.51\\ -4.51\\ -3.85\\ -4.71\end{tabular} & \begin{tabular}[c]{@{}l@{}}0.79\\ 0.79\\ 0.75\\ 0.79\end{tabular} \\ \hline
1/32       & \begin{tabular}[c]{@{}l@{}}CsiNet\_Original\\ CsiNet\_Quantized\\ CsiNet\_Pruned\\ CsiNet\_Weight\_Clustered\end{tabular} & \begin{tabular}[c]{@{}l@{}}1.037\\ 0.281\\ 0.591\\ 0.221\end{tabular} & \begin{tabular}[c]{@{}l@{}}6238.10\\ 5799.49\\ 1764.79\\ 6137.67\end{tabular} & \begin{tabular}[c]{@{}l@{}}-6.24\\ -6.29\\ -6.31\\ -5.26\end{tabular} & \begin{tabular}[c]{@{}l@{}}0.89\\ 0.89\\ 0.89\\ 0.85\end{tabular} & \begin{tabular}[c]{@{}l@{}}-2.81\\ -2.81\\ -2.38\\ -2.81\end{tabular} & \begin{tabular}[c]{@{}l@{}}0.67\\ 0.67\\ 0.63\\ 0.66\end{tabular} \\ \hline
1/64       & \begin{tabular}[c]{@{}l@{}}CsiNet\_Original\\ CsiNet\_Quantized\\ CsiNet\_Pruned\\ CsiNet\_Weight\_Clustered\end{tabular} & \begin{tabular}[c]{@{}l@{}}0.537\\ 0.156\\ 0.303\\ 0.129\end{tabular} & \begin{tabular}[c]{@{}l@{}}5719.81\\ 5452.50\\ 1463.46\\ 5712.81\end{tabular} & \begin{tabular}[c]{@{}l@{}}-5.84\\ -5.89\\ -5.76\\ -5.42\end{tabular} & \begin{tabular}[c]{@{}l@{}}0.87\\0.87\\ 0.87\\ 0.86\end{tabular} & \begin{tabular}[c]{@{}l@{}}-1.93\\ -1.93\\ -1.62\\ -1.85\end{tabular} & \begin{tabular}[c]{@{}l@{}}0.59\\ 0.59\\ 0.55\\ 0.58\end{tabular} \\ \hline
\end{tabular}
\end{adjustbox}
\end{table*}

\begin{table*}[!htb]
\centering\small
\caption{Combined Model Compression Techniques on CsiNet}
\label{table:CsiNetResults2}
\begin{adjustbox}{width=0.9\textwidth}
\begin{tabular}{|l|l|l|l|l|l|l|l|}
\hline
\textbf{\(\gamma\)} & \textbf{Model}                                                                                & \textbf{Size (MB)}                                & \textbf{Inference (\(\mu\)s)}                            & \textbf{Indoor NMSE}                              & \textbf{Indoor \(\rho\)}                                 & \textbf{Outdoor NMSE}                             & \textbf{Outdoor \(\rho\)}                                \\ \hline
1/4        & \begin{tabular}[c]{@{}l@{}}CsiNet\_Quantized\_Pruned\\ CsiNet\_Quantized\_Clustered\end{tabular} & \begin{tabular}[c]{@{}l@{}}0.942\\ 1.085\end{tabular} & \begin{tabular}[c]{@{}l@{}}2069.26\\ 6501.77\end{tabular} & \begin{tabular}[c]{@{}l@{}}-17.43\\ -11.69\end{tabular} & \begin{tabular}[c]{@{}l@{}}0.99\\ 0.96\end{tabular} & \begin{tabular}[c]{@{}l@{}}-8.70\\ -9.82\end{tabular} & \begin{tabular}[c]{@{}l@{}}0.91\\ 0.93\end{tabular} \\ \hline
1/16       & \begin{tabular}[c]{@{}l@{}}CsiNet\_Quantized\_Pruned\\ CsiNet\_Quantized\_Clustered\end{tabular} & \begin{tabular}[c]{@{}l@{}}0.256\\ 0.272\end{tabular} & \begin{tabular}[c]{@{}l@{}}1468.55\\ 5921.54\end{tabular} & \begin{tabular}[c]{@{}l@{}}-8.67\\ -7.39\end{tabular} & \begin{tabular}[c]{@{}l@{}}0.93\\ 0.91\end{tabular} & \begin{tabular}[c]{@{}l@{}}-3.85\\ -4.71\end{tabular} & \begin{tabular}[c]{@{}l@{}}0.75\\ 0.79\end{tabular} \\ \hline
1/32       & \begin{tabular}[c]{@{}l@{}}CsiNet\_Quantized\_Pruned\\ CsiNet\_Quantized\_Clustered\end{tabular} & \begin{tabular}[c]{@{}l@{}}0.146\\ 0.166\end{tabular} & \begin{tabular}[c]{@{}l@{}}1416.78\\ 5734.94\end{tabular} & \begin{tabular}[c]{@{}l@{}}-6.32\\ -5.26\end{tabular} & \begin{tabular}[c]{@{}l@{}}0.88\\ 0.85\end{tabular} & \begin{tabular}[c]{@{}l@{}}-2.38\\ -2.81\end{tabular} & \begin{tabular}[c]{@{}l@{}}0.63\\ 0.66\end{tabular} \\ \hline
1/64       & \begin{tabular}[c]{@{}l@{}}CsiNet\_Quantized\_Pruned\\ CsiNet\_Quantized\_Clustered\end{tabular} & \begin{tabular}[c]{@{}l@{}}0.085\\ 0.093\end{tabular} & \begin{tabular}[c]{@{}l@{}}1395.28\\ 5432.54\end{tabular} & \begin{tabular}[c]{@{}l@{}}-5.73\\ -5.42\end{tabular} & \begin{tabular}[c]{@{}l@{}}0.87\\ 0.86\end{tabular} & \begin{tabular}[c]{@{}l@{}}-1.62\\ -1.85\end{tabular} & \begin{tabular}[c]{@{}l@{}}0.55\\ 0.58\end{tabular} \\ \hline
\end{tabular}
\end{adjustbox}
\end{table*}

\section{Numerical Results and Analysis}\label{results}
In this section we first evaluate and explain the results observed from applying pruning, post-training quantization and weight clustering independently to CsiNet. These results are summarized in Table ~\ref{table:CsiNetResults1}. We then present the results obtained from combining different model compression techniques together. These results are summarized in Table ~\ref{table:CsiNetResults2}. Next, we investigate the effects of applying different levels of model sparsity percentages and quantization levels - both independently and jointly. These results presented in Figure ~\ref{fig:PruneQuant} provide insights on the trade-offs between accuracy, inference time, and model size - and indicate the sparsity and quantization levels that provide an excellent operating point for CsiNet. Finally, we demonstrate the need to leverage and implement sparse inference acceleration for hardware implementations of pruned neural networks. These results are presented in Figure ~\ref{XNNPack} and show the gains in inference time with the XNNPack sparse inference accelerator \cite{b15} \cite{b16}.

\subsection{Pruning}
The pruned models used to collect the results in Table ~\ref{table:CsiNetResults1} and~\ref{table:CsiNetResults2} have 50\% sparsity in the dense layers. This value was experimentally chosen to give a good trade-off between model size, inference time and NMSE (results shown in Sec V-F).

\subsubsection{\textbf{Model Size}} Pruning successfully reduced the size of CsiNet for the four different compression ratios \(\gamma\) used in \cite{b6}. An average size reduction of 43.9\% was achieved for the four different \(\gamma\) values.

\subsubsection{\textbf{Inference}} Pruning achieved the lowest latency compared to the original, quantized and weight clustered models for all values of \(\gamma\). On average, pruning provided 66.8\% decrease in inference time when compared to the original CsiNet model. This is because 0 weights were skipped during inference when XNNPack acceleration is implemented. These results demonstrate the effectiveness of pruning in enabling low-latency in wireless communications.

\subsubsection{\textbf{NMSE and \(\rho\)}} The performance of the pruned CsiNet models was comparable to the original models for most cases. In an indoor environment, \(\rho\) for the original and pruned models were identical for all values of \(\gamma\). For \(\gamma\) = 1/4, 1/16 and 1/32 in an indoor environment the pruned model outperformed the original model in terms of NMSE by a small margin. In an outdoor environment the original model outperformed the pruned model in \(\rho\) and NMSE. This was most significant for \(\gamma\) = 1/16 where the NMSE of the pruned model was on average 15.3\% larger than the original model and suggests that for compressed CsiNets in an outdoor environment a lower sparsity level is needed to maintain an on par accuracy.

\subsection{Post-Training Quantization}
Dynamic range quantization was used to collect the results in Table ~\ref{table:CsiNetResults1} and Table ~\ref{table:CsiNetResults2}. This level of quantization was experimentally chosen to give the best trade-off between model size, inference time and accuracy (results shown in Sec V-F).

\subsubsection{\textbf{Model Size}} Post-training dynamic range quantization was very effective in reducing the size of CsiNet for all compression ratios \(\gamma\) - reducing the size of CsiNet by 73.1\% on average. This suggests that post-training quantization is an effective and quick method to apply to wireless deep neural networks as the network size is dominated by its weights and activations which can be significantly reduced by quantization with minimal impact to performance.

\subsubsection{\textbf{Inference}}
Applying post-training quantization reduced inference compared to the original CsiNet model for all values of \(\gamma\). The inference time for the quantized model was on average 10.4\% less than the original model. This acceleration is because computations performed with 8-bit weights and activations are less computationally intensive compared to 32 bit floating point computations. Post-training dynamic range quantization was most effective for \(\gamma\) = 1/4, resulting in 22.6\% decrease in inference time. This is because for larger models there are more weights to quantize which leads to more gains.

\subsubsection{\textbf{NMSE and \(\rho\)}}
Applying post-training dynamic range quantization to the original CsiNet models had very little impact on the values of NMSE and \(\rho\). For all values of \(\gamma\) in an indoor and outdoor environment \(\rho\) was equivalent between the original CsiNet models and the quantized models. In an indoor environment for \(\gamma\) = 1/32, 1/64 the quantized model outperformed the original model in terms of NMSE by a very small margin and for \(\gamma\) = 1/4 the original model outperformed the quantized model by a negligible amount. These results suggest that post-training dynamic range quantization has a very minimal effect on model performance, and in essence the model size and latency gains can be ``free".

\begin{figure}
\centering
		\subfigure[]{\includegraphics[height=0.22\textwidth, width=0.42\textwidth]{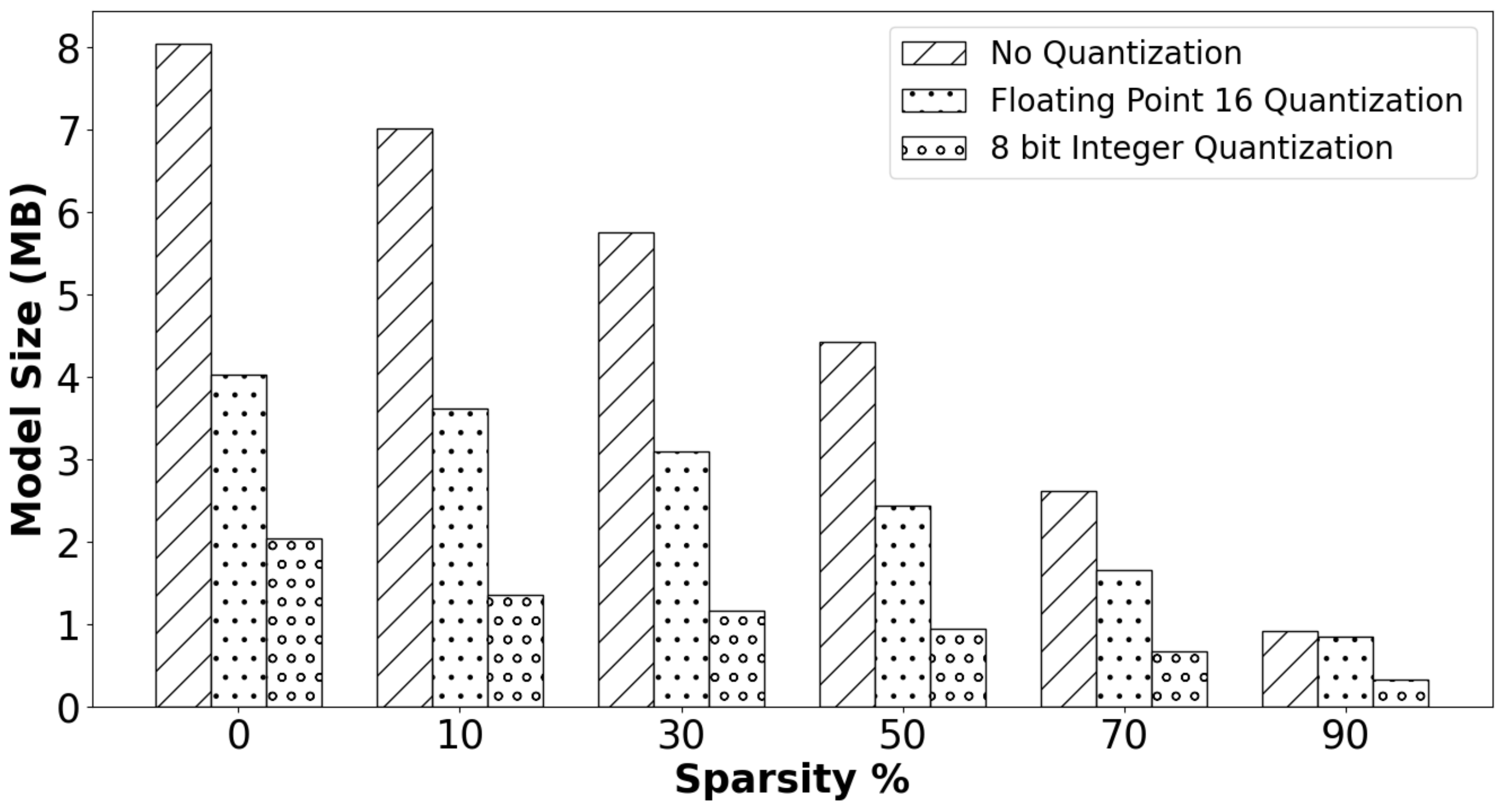}\label{a}}
		\subfigure[]{\includegraphics[height=0.22\textwidth, width=0.42\textwidth]{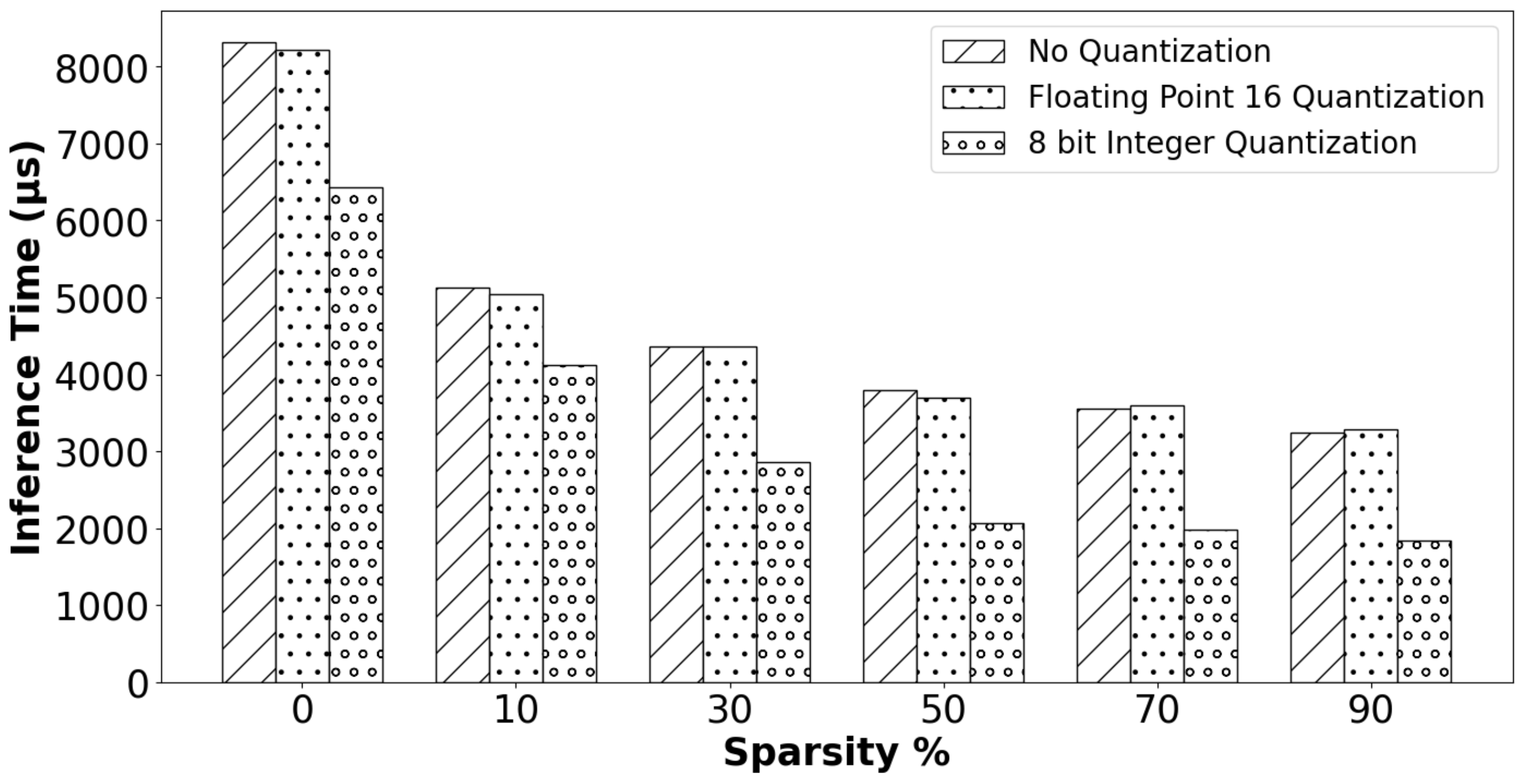}\label{b}}
		\subfigure[]{\includegraphics[keepaspectratio, width=0.42\textwidth]{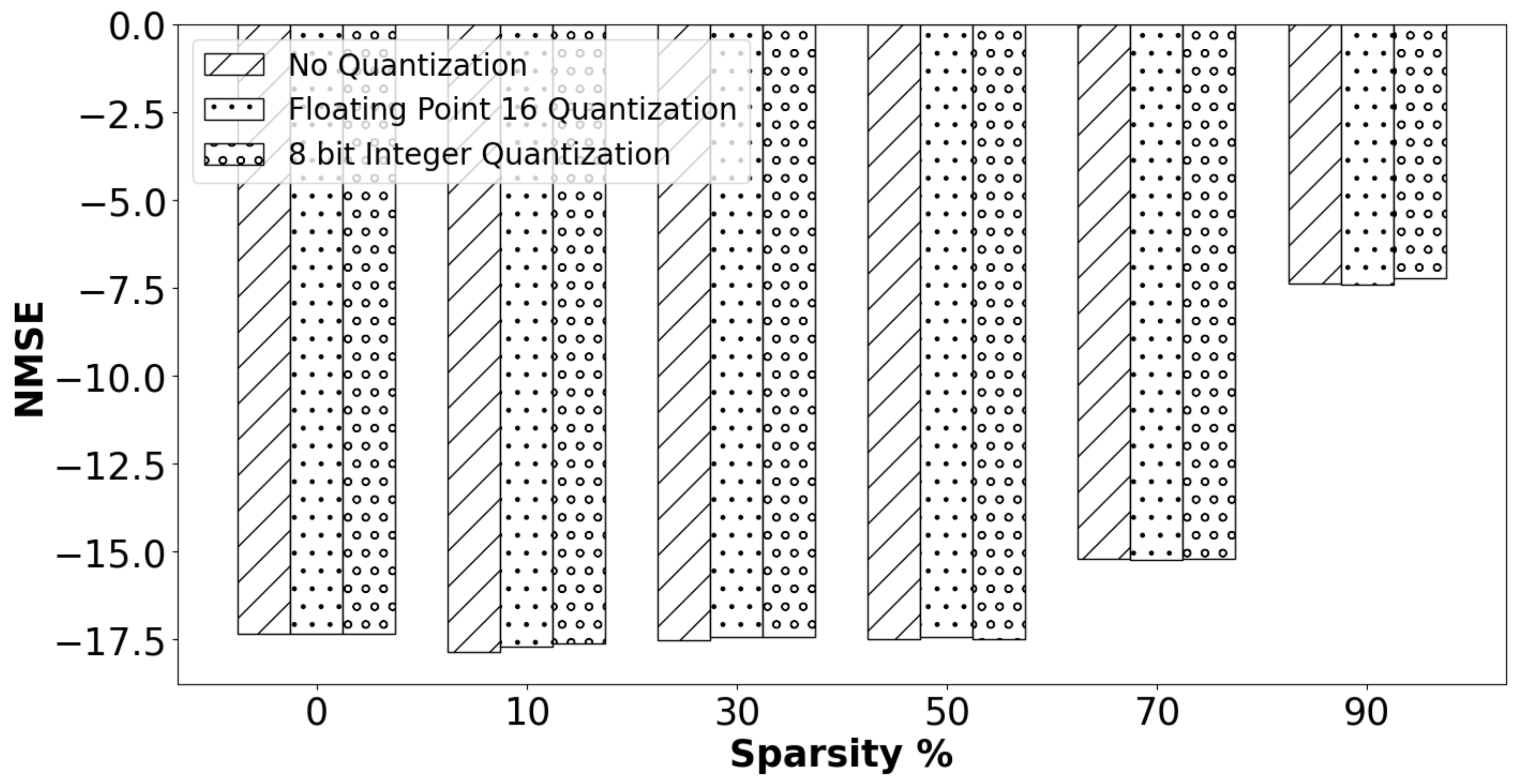}\label{c}}
        \caption{Results of Varying Sparsity and Quantization Levels for \(\gamma\) = 1/4 in an indoor environment (a) Model Size (b) Inference Time (c) NMSE}
        \label{fig:PruneQuant}
\end{figure}

\subsection{Weight Clustering}
The weight clustered models used to collect the results in Table ~\ref{table:CsiNetResults1} and Table ~\ref{table:CsiNetResults2} use kmeans++ centroid initialization and 32 clusters per network layer. This value was experimentally chosen to give a good trade-off between model size, inference time and accuracy.

\subsubsection{\textbf{Model Size}} Weight clustering had the best performance in terms of model size reduction compared to pruning and quantization. Weight clustering reduced the size of CsiNet by 79.2\% on average for the four different values of \(\gamma\). This suggests that weight clustering is an effective method to apply to wireless models to achieve significant reduction in model size. Weight clustering works by grouping weights into N clusters. Therefore by choosing a suitable value of N for any wireless model, significant model size reductions are expected.

\subsubsection{\textbf{Inference}} 
Weight clustering did not provide inference time gains. As shown in Table ~\ref{table:CsiNetResults1}, the weight clustered models had a comparable inference time to that of the base models. This is expected since weight clustering reduces the number of unique weights but does not affect the total number of inference operations.

\subsubsection{\textbf{NMSE and \(\rho\)}}
Weight clustering performed very well in the outdoor environment but had a worse performance in the indoor environment compared to pruning and quantization. For the indoor environment, \(\rho\) was on average 2.71\% smaller compared to the original models for all values of \(\gamma\). The NMSE for the weight clustered models was on average 17.5\% larger compared to the original CsiNet models. The weight clustered model performed much better in an outdoor environment and provided the best NMSE out of all methods for \(\gamma=4\). This suggests that extra care is needed when applying weight clustering to different models and environments and a generalized approach is not as reliable as with pruning and quantization.

\subsection{Pruning and Post-Training Quantization}

\subsubsection{\textbf{Model Size}} Combining pruning and post-training quantization achieved the smallest model size compared to all other results presented in this work. Pruning and post-training quantization reduced the size of CsiNet by 86.5\% on average for the four different values of \(\gamma\). This result is higher than results achieved by pruning and quantization individually, which demonstrates that combining pruning and quantization is an effective way to significantly reduce the model size.

\subsubsection{\textbf{Inference}} Combining pruning and post-training dynamic range quantization resulted in the lowest latency out of all the models presented in Table ~\ref{table:CsiNetResults1} and ~\ref{table:CsiNetResults2}. The inference time for the pruned dynamic range quantized models were on average 76.2\% shorter than the original CsiNet models. These results demonstrate the effectiveness of pruning and post training dynamic range quantization in enabling low-latency in wireless communications. 

\subsubsection{\textbf{NMSE and \(\rho\)}} The results for NMSE and the cosine similarity \(\rho\) when combining pruning and post-training dynamic range quantization were comparable to the results obtained by the original model for an indoor environment. The model unperformed slightly in an outdoor environment which is expected because the original pruned model before quantization also underperformed in an outdoor environment.

\subsection{Weight Clustering and Post-Training Quantization}

The weight clustered models used to collect the results in Table ~\ref{table:CsiNetResults1} and Table ~\ref{table:CsiNetResults2} use 32 clusters and kmeans++ centroid initialization. These clustering parameters were experimentally chosen to give high accuracy and low model size. 

\subsubsection{\textbf{Model Size}} Combining weight clustering and post-training dynamic range quantization resulted in a large reduction in model size compared to the original CsiNet models. Weight clustering and quantization reduced the size of CsiNet by 85.0\% on average for the four different values of \(\gamma\). This reduction is higher than results achieved by weight clustering and quantization individually, this is because once the weights are grouped into N clusters they are then represented as 8 bit integers instead of 32 bit floating point values. However, because there are less unique weight values, there are less weights to quantize and we only see a small improvement when we combine the two techniques.

\subsubsection{\textbf{Inference}}
The combination of weight clustering and post-training dynamic range quantization resulted in lower inference times compared to the inference times of the original CsiNet models. This is mainly due to inference improvements that result from post-training dynamic range quantization. The inference time for the weight clustered and quantized models were on average 10.5\% less than the original CsiNet models.

\subsubsection{\textbf{NMSE and \(\rho\)}}
The results for NMSE and \(\rho\) when combining weight clustering and post-training dynamic range quantization were equivalent to the results achieved by weight clustering individually. As we saw earlier, post-training dynamic quantization had minimal effects on the NMSE and \(\rho\) and this is confirmed again here as the results obtained from combining both compression techniques were identical to the results obtained by the weight clustered models.

\subsection{Varying Pruning and Quantization Parameters}
\subsubsection{\textbf{Model Size}} Figure ~\ref{a} highlights the impact of model sparsity \% and quantization level on model size. It is evident that increasing model sparsity leads to a large decrease in model size regardless of the quantization level. Dynamic range quantization provides the smallest model size, followed by floating point 16 quantization and then the original model in floating point 32. At a high model sparsity \%, the effects of quantization become less noticeable. This is expected because there are less weights to quantize at a higher model sparsity. 
\subsubsection{\textbf{Inference}} Figure ~\ref{b} highlights the impact of model sparsity \% and quantization level on inference time. The higher the model sparsity \%, the lower the inference time. Inference time for floating point 16 quantization was very similar to the original floating point 32 model. Dynamic range quantization provided the best inference improvements compared to all other levels of quantization.
\subsubsection{\textbf{NMSE}} Figure ~\ref{c} highlights the impact of model sparsity \% and quantization level on the NMSE. The quantization level has minimal effects on the NMSE, and the model sparsity \% is the dominant factor. Therefore, quantization to 8-bits should always be conducted to reap its model size and acceleration gains. At 70\% sparsity we see a drop in NMSE and at 90\% sparsity we see the largest drop. These results indicate that the configuration of 50\% sparsity and 8-bit quantization provides the largest complexity reduction gains without an impact to accuracy. 
\subsection{Accelerating Sparse Models}
\begin{figure}
\centering
		\includegraphics[keepaspectratio, width=0.45\textwidth]{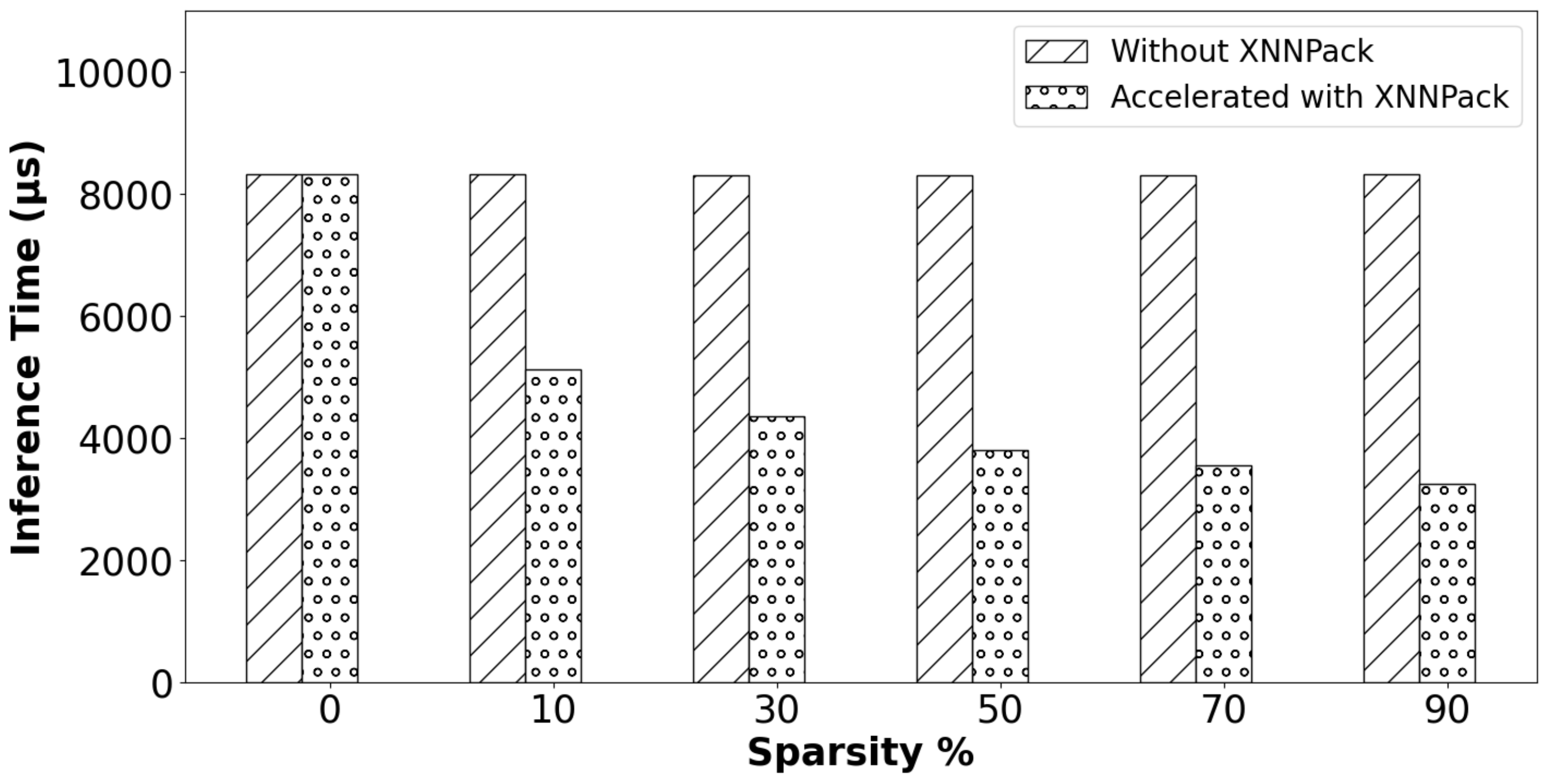}
		\caption{XNNPack Sparse Model Acceleration}
		\label{XNNPack}
\end{figure}

\textbf{Inference:} Figure ~\ref{XNNPack} demonstrates the importance of XNNPack sparse inference to improve the inference time of the resultant sparse models after pruning. Pruning provides remarkable improvements in model size, however without XNNPack sparse inference, the inference time of the pruned model is comparable to the original model. This highlights the importance of optimized neural network inference operators, and opens the door for future research that can focus on accelerating sparse models further \cite{b15} \cite{b16}.

\section{Conclusion}\label{conclusion}
This paper presented a comprehensive study on applying three common model compression techniques to CsiNet, a wireless communications deep learning model for massive MIMO CSI feedback. We demonstrated the effectiveness of these techniques in terms of model size, inference time, and accuracy - and analyzed the interaction of combining multiple compression techniques to obtain even more efficient networks that do not impact performance. Furthermore, we show the importance of accelerating sparse models to achieve low inference times on commodity hardware. 
The results of this paper demonstrate the importance of adopting these techniques when developing deep learning wireless models to obtain tiny, efficient, and accurate models that can provide low-latency processing and consume limited memory and power. These methods are crucial to pave the way for practical adoption and deployments of deep learning-based techniques in commercial wireless systems. 



\end{document}